\documentclass[letterpaper,aps, prb, showpacs, twocolumn, 
amssymb,superscriptaddress]{revtex4-1}
\usepackage[dvipdfmx]{graphicx}
\usepackage{bm, amsmath, amsfonts}
\usepackage{color}
\usepackage{subfigure}
\usepackage{graphicx}

\usepackage{hyperref}
\usepackage{tikz}
\usepackage{pgfplots}
\pgfplotsset{compat=1.12}
\pgfplotsset{compat=newest}
\usepgfplotslibrary{fillbetween,patchplots}    
\usetikzlibrary{shapes.misc,patterns}
\usepackage{braket}
\usepackage{mathtools}
\usepackage{commath}

\def\dagg{^\dagger}
\def\su{\uparrow}

\def\vac{\ket{\textup{vac}}}

\newcommand{\new}[1]{#1}                    

\begin{document}


\title{
Exact ground states for interacting Kitaev chains 
}

\author{Jurriaan Wouters}
\affiliation{
Institute for Theoretical Physics, Center for Extreme Matter and Emergent Phenomena,
Utrecht University, Princetonplein 5, 3584 CE Utrecht, The Netherlands
}

\author{Hosho Katsura}
\affiliation{
Department of Physics, Graduate School of Science, The University of Tokyo, 
Hongo, Tokyo 113-0033
}

\author{Dirk Schuricht}
\affiliation{
Institute for Theoretical Physics, Center for Extreme Matter and Emergent Phenomena,
Utrecht University, Princetonplein 5, 3584 CE Utrecht, The Netherlands
}

\date{\today}
\begin{abstract}
We introduce a frustration-free, one-dimensional model of spinless fermions with hopping, p-wave superconducting pairing and alternating chemical potentials. The model possesses two exactly degenerate ground states even for finite system sizes. We present analytical results for the strong Majorana zero modes, the phase diagram and the topological order. Furthermore, we generalise our results to include interactions.
\end{abstract}


\maketitle

\section{Introduction}\label{sec:intro}
Majorana fermions have attracted a lot of attention over the last two decades. Motivated by their anticipated future role\cite{bravyi2000fermionic,alicea2011non} in quantum computing applications, systems supporting Majorana zero modes have been widely studied in condensed-matter physics, culminating in recent experiments\cite{mourik2012signatures,deng2012anomalous,das2012zero,churchill2013superconductor,nadj2014observation,deng2017majorana,lutchyn2018majorana} on superconductor-semiconductor nanowire systems.

The prime example of a model possessing Majorana zero modes is the Kitaev chain.\cite{kitaev2001unpaired} It describes spinless fermions on a tight-binding chain with open boundary conditions, which are subject to p-wave superconducting pairing with fermionic parity as symmetry operator. Depending on the parameters, the system will be either in its trivial or its topological phase. The latter is marked by a two-fold degenerate ground state, with corresponding zero-energy modes in the insulating gap. The zero modes are hermitian and normalisable, localised at the boundaries of the chain, commute with the Hamiltonian and anti-commute with fermion parity operator, making them Majorana edge zero modes.\cite{fendley2012parafermionic} Furthermore, because the two ground states live in the two different symmetry sectors, hybridisation is exponentially suppressed, hence the fermionic parity is protected by topology. Theoretical works on disorder,\cite{brouwer2011probability,lobos2012interplay,degottardi2013majorana,altland2014quantum,crepin2014nonperturbative,gergs2016topological,mcginley2017robustness,kells2018localization} dimerisation\cite{wakatsuki2014fermion,sugimoto2017reentrant,ezawa2017exact,wang2017characterization} and interactions\cite{gangadharaiah2011majorana,stoudenmire2011interaction,sela2011majorana,crepin2014nonperturbative,hassler2012strongly,thomale2013tunneling,milsted2015statistical,rahmani2015phase,katsura2015exact,chan2015multiple,sarkar2016physics,gergs2016topological,ezawa2017exact,wang2017characterization,miao2017exact,kells2018localization} have shown that the topological phase is very robust against various perturbations. Furthermore, via the non-local Jordan--Wigner transformation, the Kitaev chain can be mapped to a transverse-field Ising/XY chain, with the mentioned perturbations leading to more general spin-1/2 spin chains.

In this article we propose an inhomogeneous modification to the Kitaev chain, for which the zero modes and ground state obtain special properties. It turns out that the Majorana mode energy in this model becomes exactly zero, even for finite lattice sizes. This is in contrast with the generic Kitaev chain in its topological phase, where the energy decays exponentially with the length of the system. Moreover, we can obtain the ground states in a simple product form. The latter is equivalent to the model being frustration-free, meaning all local Hamiltonians are simultaneously minimised when projected onto the ground-state subspace. We will elaborate more on this notion later in the article. Well known frustration-free models are the AKLT chain\cite{affleck1987rigorous,affleck1988valence} and the Kitaev toric code.\cite{kitaev2003fault} There has also been progress concerning spin chains/Majorana models.\cite{asoudeh2007entangled,katsura2015exact} An overview of homogeneous frustration-free XYZ/interacting-Majorana chains was given in Ref.~\onlinecite{jevtic2017frustration}.

Furthermore, we also introduce an interacting frustration-free model. This is an extension of the non-interacting inhomogeneous Kitaev chain, obtained by exploiting the local fermion-parity invariance. For this interacting model, we derive the ground-state energy analytically and give an estimate on the spectral gap. The exact ground states are inherited from the non-interacting model, which gives the opportunity to analytically compute zero-temperature correlation functions.

This article is organised as follows: In Sec.~\ref{sec:noninteracting} we introduce the non-interacting model with alternating chemical potentials. We discuss in detail its properties, including the construction of the exact ground states, strong zero modes, phase diagram and topological order. In Sec.~\ref{sec:fullinhom} we briefly discuss the construction of exact strong zero modes in the system with completely inhomogeneous chemical potentials. Finally, in Sec.~\ref{sec:interactions} we return to the alternating setup but include interactions. The exact, two-fold degenerate ground state of the resulting model is calculated and the phase diagram is obtained, before we end with a conclusion. Technical details of our derivations are deferred to several appendices.

\section{Non-interacting model}\label{sec:noninteracting}
We begin by introducing the non-interacting model and discussing its properties in detail. Interactions will be added in Sec.~\ref{sec:interactions}. 

\subsection{Hamiltonian}
We consider an open chain of length $L$ supporting spinless fermions. The creation and annihilation operators on site $j$ are given by $c_j\dagg$ and $c_j$ respectively, satisfying canonical anti-commutation relations $\{c_i,c_j\}=\{c_i^\dagger,c_j^\dagger\}=0$ and $\{c_i,c_j^\dagger\}=\delta_{ij}$. The number operator on site $j$ is defined as $n_j=c_j\dagg c_j$. The Hamiltonian of the non-interacting model is given by
\begin{eqnarray}
H&=&-\sum_{j=1}^{L-1}\left[t(c\dagg_jc_{j+1}+c\dagg_{j+1}c_j)+\Delta(c\dagg_jc_{j+1}^\dagger+c_{j+1}c_j)\right] \nonumber\\
&&-\sum_{j=1}^{L}q_j(2n_j-1)+S,\label{eq:kitaev}\\
S&=&\frac12\left[q_1(2n_1-1)+q_L(2n_L-1)\right],
\end{eqnarray}
where $t$ and $\Delta$ are the hopping and pairing energies respectively. For later use we introduce the parametrisation 
\begin{equation}
t=\frac{\eta+\eta^{-1}}2,\quad \Delta=\frac{\eta-\eta^{-1}}2,
\label{eq:tDeltaeta}
\end{equation}
which effectively removes the overall energy scale and thus reduces the number of parameters by one. An overall scale can be reintroduced, not changing anything qualitatively in the rest of the article. Furthermore, $q_j$ denotes a chemical potential at lattice site $j$. Except for Sec.~\ref{sec:fullinhom} we will consider the alternating setup 
\begin{equation}
q_j=\begin{cases}
q,&\textup{ if $j$ odd,}\\
q^{-1},&\textup{ if $j$ even.}
\end{cases}
\label{eq:parameters}
\end{equation}
Finally we note that the surface term $S$ alters the chemical potential on the edges. This allows us to rewrite Eq.~\eqref{eq:kitaev} as sum of local Hamiltonians
\begin{eqnarray}
H&=&\sum_{j=1}^{L-1}h_j,\\
h_j&=&-t(c_j\dagg c_{j+1}+c_{j+1}\dagg c_j)-\Delta(c_j\dagg c_{j+1}\dagg+c_{j+1} c_j) \nonumber\\
&&-\frac{q_j}2(2n_j-1) -\frac{q_{j+1}}2(2n_{j+1}-1).\label{eq:lochamfermion}
\end{eqnarray}
The homogeneous system (i.e., $q_j$ constant) is equivalent to the non-interacting point of the model studied in Ref.~\onlinecite{katsura2015exact}. 

For simplicity, we will only consider even system lengths. Furthermore, we can assume $q>0$ and $\eta>0$, since charge conjugation [$c_j\rightarrow(-1)^jc\dagg_j$] is equivalent to $q\rightarrow -q$ while $\eta\rightarrow-\eta$ can be achieved by $c_j\rightarrow(-1)^jc_j$. Moreover, we can restrict $q$ and $\eta$ to be larger than 1, because  inversion ($c_j\rightarrow ic_{L-j+1}$) induces $q\rightarrow q^{-1}$ and $c_j\rightarrow ic_j$ induces $\eta\rightarrow \eta^{-1}$.

The total fermion number $F=\sum_jn_j=\sum_jc\dagg_jc_j$ is not conserved by the Hamiltonian. However, the fermionic parity, i.e., the fermion number modulo two is a symmetry of the model,
\begin{equation}
[H,(-1)^F]=0,
\end{equation}
where $(-1)^F$ is the fermionic parity operator.

\subsection{Exact ground states}\label{ssec:exactgroundstates}
In general the $h_j$'s cannot be diagonalised simultaneously, because $[h_j,h_{j+1}]\neq0$. However, progress can be made in the frustration-free case. This occurs if there exists a subspace, the ground-state space, with projector $G_0$, such that every $h_j$ is minimised when projected onto this space, i.e., $h_jG_0=\epsilon_0 G_0$ with $\epsilon_0$ being the smallest eigenvalue of $h_j$. The ground states of the full Hamiltonian $H$ minimise each $h_j$ independently and the Hamiltonian $H$ is said to be frustration-free. There are many examples of frustration-free models, possibly the most well known are the AKLT chain\cite{affleck1987rigorous,affleck1988valence} and the Kitaev toric code.\cite{kitaev2003fault} An extensive discussion on frustration-free system was recently given by Jevtic and Barnett.\cite{jevtic2017frustration} They give a systematic derivation of spin chain models with a factorised ground state, which support Majorana zero modes.

In this section we will show that the model \eqref{eq:kitaev} is frustration-free. The determination of the ground-state subspace requires to minimise every local Hamiltonian, so we start by considering the two-site problem. The four states for this subsystem are $\ket{\circ\circ}=\vac$, $\ket{\bullet\circ}=c\dagg_j\vac$, $\ket{\circ\bullet}=c\dagg_{j+1}\vac$ and $\ket{\bullet\bullet}=c\dagg_jc\dagg_{j+1}\vac$, where $\vac$ denotes the vacuum state on the lattice sites $j$ and $j+1$. The local Hamiltonians also preserve fermionic parity, so we can split system in an even and an odd sector,
\begin{align}
h_j^e&=-\frac12\begin{pmatrix}
q_j+q_j^{-1}&\eta-\eta^{-1}\\
\eta-\eta^{-1}&-q_j-q_j^{-1}
\end{pmatrix},
\label{eq:hloceven}
\\
h_j^o&=-\frac12\begin{pmatrix}
q_j-q_j^{-1}&\eta+\eta^{-1}\\
\eta+\eta^{-1}&-q_j+q_j^{-1}
\end{pmatrix},
\label{eq:hlocodd}
\end{align}
which act on the basis $\{\ket{\bullet\bullet},\ket{\circ\circ}\}$ and $\{\ket{\bullet\circ},\ket{\circ\bullet}\}$ respectively. For both sectors the eigenvalues are given by $\epsilon_\pm=\pm\frac12\mathcal{N}$, with $\mathcal{N}=\sqrt{q^2+q^{-2}+\eta^2+\eta^{-2}}$. The corresponding eigenstates with energy $\epsilon_-$ are
\begin{align}
\ket{\psi_j^e}&=\left(\frac{\mathcal{N}+q_j+q_j^{-1}}{\eta-\eta^{-1}}\right)\ket{\bullet\bullet}+\ket{\circ\circ},\\
\ket{\psi_j^o}&=\left(\frac{\mathcal{N}+q_j-q_j^{-1}}{\eta+\eta^{-1}}\right)\ket{\bullet\circ}+\ket{\circ\bullet}.
\end{align}
We note that any linear combination of these eigenstates minimises the local Hamiltonian $h_j$. 

In order to find the ground state of the full system we first look for linear combinations of the eigenstates $\ket{\psi_j^{e,o}}$  that can be written as a product of single-site states. Making the ansatz
\begin{align}
\ket{\psi_{2k-1}^e}\pm x_1\ket{\psi_{2k-1}^o}&=(x_0c\dagg_{2k-1}\pm 1)(x_1c\dagg_{2k}\pm 1)\vac,\\
\ket{\psi_{2k}^e}\pm x_0\ket{\psi_{2k}^o}&=(x_1c\dagg_{2k}\pm 1)(x_0c\dagg_{2k+1}\pm 1)\vac,
\end{align} 
where we distinguish between even and odd sites because of the alternating nature of the model, the coefficients are found to be
\begin{align}
(x_0)^2&=\frac{\mathcal{N}+q+q^{-1}}{\eta+\eta^{-1}}\frac{\mathcal{N}+q-q^{-1}}{\eta-\eta^{-1}},\label{eq:x0}\\
(x_1)^2&=\frac{\eta+\eta^{-1}}{\eta-\eta^{-1}}\frac{\mathcal{N}+q+q^{-1}}{\mathcal{N}+q-q^{-1}}.\label{eq:x1}
\end{align}
Now, since $h_j$ commutes with $x_{0,1}c\dagg_k\pm1$ for $k\neq j,j+1$, the ground states minimising all local Hamiltonians and consequently the full system  are
\begin{equation}\label{eq:factorizedGS}
\ket{\Psi^\pm}=\prod_{k=1}^{L/2}(x_0 c_{2k-1}\dagg\pm 1)(x_1 c_{2k}\dagg \pm1)\vac
\end{equation}
with ground-state energy 
\begin{equation}
E_0=-\frac{L-1}2\mathcal{N}.
\end{equation}
Note that for $\eta=1$ the ground state becomes non-degenerate, as both $x_0$ and $x_1$ diverge and the ground state is a fully filled state. In Sec.~\ref{sec:MZM} we will see that the system possesses strong zero modes supporting the two-fold degeneracy in the ground state. Moreover, the zero modes indicate that there are no more linearly independent ground states, i.e., the ground-state subspace is two-dimensional.

At this point we have to note that the model in question is non-interacting, hence finding the ground state is always possible. However, this specific factorised form is reserved for only a small class of models. An alternative, more direct approach for finding the ground states is discussed in App.~\ref{app:lindblad}. There the notion of Lindblad operators is employed to verify the states in Eq.~\eqref{eq:factorizedGS} span the full ground-state subspace.

As noted above, the Hamiltonian \eqref{eq:kitaev} preserves fermionic parity. However, $\ket{\Psi^\pm}$ belong neither to the even nor the odd parity sector. More explicitly, because $(-1)^F(x_{0,1}c\dagg_k\pm1)=-(x_{0,1}c\dagg_k\mp1)(-1)^F$, the action of the fermionic parity on the ground states is $(-1)^F\ket{\Psi^\pm}=\ket{\Psi^\mp}$. Also, the states are not orthogonal. However, the linear combinations
\begin{align}
\ket{\Psi^e}&=\frac1{\sqrt{2(N+M)}}\left(\ket{\Psi^+}+\ket{\Psi^-}\right),\label{eq:GSnorme}\\
\ket{\Psi^o}&=\frac1{\sqrt{2(N-M)}}\left(\ket{\Psi^+}-\ket{\Psi^-}\right),\label{eq:GSnormo}
\end{align} 
belong to the even and odd sector, respectively, and thus are orthogonal to each other. The  coefficients that normalise the states are 
\begin{align}
N&:=\braket{\Psi^\pm|\Psi^\pm}=\left[(x_0^2+1)(x_1^2+1)\right]^{L/2},\label{eq:N}\\
M&:=\braket{\Psi^\mp|\Psi^\pm}=\left[(x_0^2-1)(x_1^2-1)\right]^{L/2},\label{eq:M}
\end{align}
which satisfy $N/M=\eta^L$.

The ground states \eqref{eq:GSnorme} and \eqref{eq:GSnormo} are locally indistinguishable, as can be seen as follows:\cite{katsura2015exact} Consider any local operators with an even ($O_e$) and odd ($O_o$) number of fermion creation an annihilation operators, supported on a sublattice $1<j_1<\ldots<j_k<L$, such that $j_k-j_1=\ell-1$. Immediately, we note that $\braket{\Psi^e|O_o|\Psi^e}=\braket{\Psi^o|O_o|\Psi^o}=0$, since $O_o$ changes the fermionic parity of the states, while $\ket{\Psi^{e,o}}$ belong respectively to the even or odd sector. Furthermore, the even local operators we can bound as (see App.~\ref{app:correlators})
\begin{equation}\label{eq:difference}
|\braket{\Psi^e|O_e|\Psi^e}-\braket{\Psi^o|O_e|\Psi^o}|\leq K\norm{O_e}e^{-L/\xi},
\end{equation}
where $\norm{O_e}$ is the operator norm [as defined in Eq.~\eqref{eq:operatornorm}], $K$ is a constant, small compared to $e^{L/\xi}$, and
\begin{equation}
\xi=\frac1{\ln[\max(\eta,\eta^{-1})]}>0
\end{equation}
is the correlation length. Hence it is not possible to distinguish the two ground states \eqref{eq:GSnorme} and \eqref{eq:GSnormo} by measuring local expectation values in a thermodynamically large system.

Finally let us rewrite the system \eqref{eq:kitaev} in the form of a spin chain using the Jordan--Wigner transformation
\begin{equation}\label{eq:jordanwigner}
\begin{aligned}
c_j&=\prod_{k=1}^{j-1}\left(-\sigma_k^z\right)\frac{\sigma^x_j- i\sigma^y_j}2,\\
c_j\dagg&=\prod_{k=1}^{j-1}\left(-\sigma_k^z\right)\frac{\sigma^x_j+ i\sigma^y_j}2,
\end{aligned}
\end{equation}
with $\sigma^i$ for $i=x,y,z$ denoting the Pauli matrices. Plugging this in into Eq.~\eqref{eq:lochamfermion} results in a XY chain in a transverse field, 
\begin{equation}\label{eq:lochamspin}
h_j=-\frac12(\eta\sigma_j^x\sigma_{j+1}^x +\eta^{-1}\sigma_j^y\sigma_{j+1}^y +q_j\sigma_j^z+q_j^{-1}\sigma_{j+1}^z).
\end{equation}
Now it can be confirmed along the lines of Refs.~\onlinecite{muller1985implications} and~\onlinecite{KURMANN1982235} that the system is frustration-free for all values of $\eta$ and $q$. 

\subsection{Strong zero modes}\label{sec:MZM}
As we saw in the previous section, the model has a two-fold degenerate ground state. This is reminiscent of the two ground states of the Kitaev chain in its topological phase. The major difference is that due to the specific tuning of the edge chemical potential the ground states of the model \eqref{eq:kitaev} are perfectly degenerate and uncoupled for all system sizes.  On the other hand, in a generic non-interacting Kitaev model, the coupling between the ground states decays exponentially with the length of the system.\cite{kitaev2001unpaired}

The perfect degeneracy of the ground states in the system \eqref{eq:kitaev} suggests that there exists a single-particle mode $T_0$ with zero energy, mapping one ground state to the other, i.e., $\ket{\Psi^o}\propto T_0\ket{\Psi^e}$. This mode must anti-commute with the fermionic parity operator, ($\{(-1)^F,T_0\}=0$), mapping one parity sector to the other. Also it has to commute at least with the ground-state part of the Hamiltonian. This is what is sometimes called a ``weak'' zero mode.\cite{alicea2012new}

In fact, since we are dealing with a non-interacting problem, the last property can be extended to the full Hilbert space, i.e., the zero mode commutes with the full Hamiltotinan, $[H,T_0]=0$. Furthermore, due to particle-hole symmetry these zero modes are necessarily Majorana modes ($T_0\dagg=T_0$). In this section we will derive explicit expressions for these modes, called \textsl{strong Majorana zero modes}, satisfying the following properties:\cite{fendley2012parafermionic}
\begin{enumerate}
	\item $T_0\dagg=T_0$,
	\item $\{(-1)^F,T_0\}=0$,
	\item $[H,T_0]=0$,
	\item $T_0^\dagger T_0=T_0^2=1$.
\end{enumerate}

To do so, we first split the spinless complex fermions into two Majorana fermions $a_j=a_j^\dagger$ and $b_j=b_j^\dagger$ per lattice site in the usual fashion,
\begin{equation}\label{eq:majorana}
c_j=\frac{a_j-i b_j}2,\quad c_j\dagg=\frac{a_j+i b_j}2.
\end{equation}
The particular choice of the hopping and superconducting parameters becomes clear in the Majorana representation, in which the local Hamiltonian becomes
\begin{equation}\label{eq:lochammaj}
h_j=-\frac{i}2(\eta b_j a_{j+1} -\eta^{-1}a_j b_{j+1} -q_j a_j b_j-q_j^{-1}a_{j+1} b_{j+1}).
\end{equation}
Requiring condition 3, or equivalently $[h_j,T_0]=0$ for all $j$, we find two zero modes of the form
\begin{align}
T^a_0&=\alpha\sum_{j=1}^{L/2}\frac{1}{\eta^{2(j-1)}}\left(a_{2j-1}-\frac{q}{\eta} a_{2j}\right),\label{eq:ZMa}\\ T^b_0&=\beta\sum_{j=1}^{L/2}\eta^{2(j-1)}\left(b_{2j-1}-q\eta b_{2j}\right)\label{eq:ZMb},
\end{align}
where $\alpha$ and $\beta$ are normalisation factors fixed by condition 4. Note that by construction they also satisfy the first two requirements above. Thus $T_0^a$ and $T_0^b$ are indeed strong zero modes. We stress that these zero modes are exact even for finite system sizes, in contrast to the modes in a generic non-interacting Kitaev chain.

Let us also comment on the localisation of the zero modes $T^a_0$ and $T^b_0$. Obviously they  decay with $\eta$ and $\eta^{-1}$ respectively, localising them at one of the edges.  For $\eta>1$,  $T^a_0$ is localised at the left ($j=1$) boundary, while $T^b_0$ lives at the right edge ($j=L$). For $\eta<1$ the situation is reversed. Thus for all $\eta\neq 1$ we have two strong zero modes localised at the opposite boundaries. At $\eta=1$ both modes are delocalised and the ground state becomes singly degenerate. 

Finally, in App.~\ref{app:actionzeromodes} we discuss the action of the zero modes on the ground states. As one would expect the zero modes map one ground state to the other, i.e., $T^{a,b}_0\ket{\Psi^e}=\ket{\Psi^o}$. This confirms that the ground states are in a different fermion parity sector, since the Majorana zero modes change the parity by 1. 

\subsection{Phase Diagram}
In the previous section we derived the existence of exact strong zero modes, supported by the two-fold degeneracy of the ground state. This hints towards a topological superconductor region in the phase diagram. Also there appears to be a phase transition at $\eta=1$. In this section we will derive the spectrum to confirm this picture by examining the gap throughout the phase diagram. We have to note that, even though we are interested in finite-size systems allowing for the presence of edge effects, here we will be considering the gap in the thermodynamic limit, as only in this limit the system can become truly gapless. In this section we are interested in the bulk gap and bulk phase transition.

\begin{figure}[t]
	\includegraphics{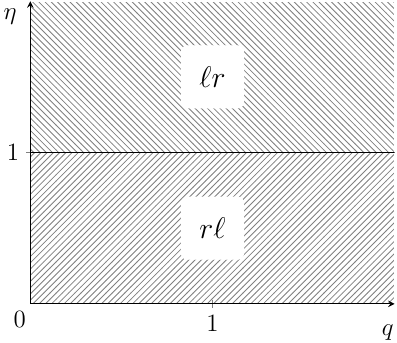}
	\caption{Phase diagram for the non-interacting model in $q-\eta$ plane. The localisation of the zero modes is indicated in the white squares, where $\ell$ and $r$ refer to the left (near site 1) and right (near site $L$) edges. The first letter denotes the location of $T^a_0$, the second of $T^b_0$. In Fig.~\ref{fig:connectiontoKitaev} a cut along $q=1$ (the homogeneous system) is shown in the phase space of the Kitaev parameters ($\mu, \Delta$). }\label{fig:criticallines}
\end{figure}

\begin{figure}
	\includegraphics{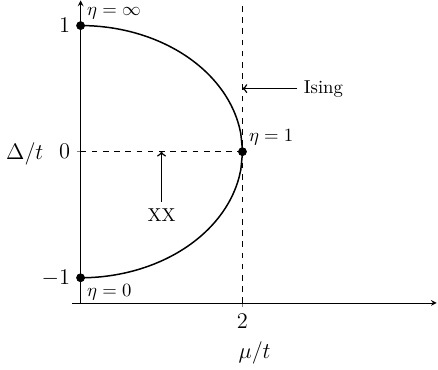}
	\caption{Cut along $q=1$ in Fig.~\ref{fig:criticallines} embedded in the $\Delta/t$-$\mu/t$ phase space for the conventional Kitaev chain. The dashed lines represent the Ising and XX phase transitions. The solid line is the cut along $q=1$. At the phase transition ($\eta=1$) the cut passes through the intersection of the Ising and XX transition lines, where the spectrum becomes quadratic at small momenta. Adapted from Ref.~\onlinecite{cherng2006entropy}. }\label{fig:connectiontoKitaev}
\end{figure}

With the local Hamiltonian in the Majorana language \eqref{eq:lochammaj} and recalling the alternating chemical potential \eqref{eq:parameters}, the Hamiltonian can be brought into matrix form 
\begin{align}
H&=\frac{i}2\sum_{j,k=1}^La_jB_{jk}b_k,\\
 B&=\begin{pmatrix}
q & \eta^{-1} & &&\\
\eta&2q^{-1} & \eta^{-1}  &&\\
&\ddots&\ddots & \ddots  &\\
&&\eta& 2q  &\eta^{-1}\\
&&& \eta  &q^{-1}
\end{pmatrix}.
\end{align}
The $L\times L$ matrix $B$ is non-hermitian, and is not necessarily diagonalisable. However, $BB^\intercal$ is hermitian and the eigenvalues are $(2\epsilon_k)^2$, with $\epsilon_k$ the single-particle energies of the model. Diagonalising the pentadiagonal $BB^\intercal$ is quite troublesome. Fortunately, we can construct a symmetric tridiagonal matrix $C$, such that $BB^\intercal=C^2$, with 
\new{
\begin{equation}\label{eq:C}
C=\begin{pmatrix}
\gamma_1 & \delta^{-1} & &&&\\
\delta^{-1}&\gamma_- & \delta  &&&\\
&\delta&\gamma_+& \delta^{-1}&&\\
&&\ddots& \ddots  &\ddots&\\
&&&\delta&\gamma_+&\delta^{-1}\\
&&&&\delta^{-1}&\gamma_L
\end{pmatrix},
\end{equation}
where
\begin{equation}
\begin{aligned}
\gamma_\pm&=(\eta^2+\eta^{-2}+2q^{\pm 2})/\mathcal{N},\\
\gamma_1&=(q^2+\eta^{-2})/\mathcal{N},\quad
\gamma_L=(q^{-2}+\eta^2)/\mathcal{N},\\
\delta&=(q\eta^{-1}+q^{-1}\eta)/\mathcal{N}.
\end{aligned}
\end{equation}}
The matrix $C$ can be diagonalised analytically, see App.~\ref{app:spectrumnonint}.
For $k=\frac{2\pi n}{L}$ with $n=1,2,\ldots,\frac{L}2-1$ we find the eigenvalues
\begin{equation}
\epsilon_k^\pm=\frac12\left(\mathcal{N}\pm\sqrt{q^2+q^{-2}+2\cos{k}}\right).
\label{eq:epsilonk}
\end{equation}
Furthermore, there are two additional modes with energies
\begin{equation}
\epsilon_0^-=0,\quad \epsilon_0^+=\frac12\mathcal{N}.
\end{equation}
Note that all eigenvalues are non-negative, because $BB^\intercal$ is positive semidefinite. The smallest non-zero eigenvalue is
\begin{equation}
\epsilon^-_{k=2\pi/L}=\frac12\left(\mathcal{N}-(q+q^{-1})+\frac{k^2}{2(q+q^{-1})}\right)+\mathcal{O}(k^4),
\end{equation}
which gives a spectral gap in the thermodynamic limit ($L\to\infty$) of
\begin{equation}\label{eq:noningap}
\Delta E=\frac{1}{2}\left[\mathcal{N}-(q+q^{-1})\right].
\end{equation}
From Eq.~\eqref{eq:noningap} we see that the gap only vanishes at $\eta=1$ indicating the phase transition. This is depicted in the $q$-$\eta$ phase diagram in Fig.~\ref{fig:criticallines}. At the phase transition the low-energy spectrum is quadratic in $k$, putting the critical model out of reach for conformal field theories. To clarify the nature of the phase transition, we consider Fig.~\ref{fig:connectiontoKitaev}. In this figure we show a cut of Fig.~\ref{fig:criticallines} along $q=1$ (the homogeneous point) projected onto the $\mu$-$\Delta$ space, i.e., using the conventional Kitaev parameters.\cite{cherng2006entropy} The dashed lines represent the Ising and XX transitions. The solid line depicts the cut along $q=1$. Clearly, the phase transition at $\eta=1$ occurs at the crossing of the Ising and XX line. As we discuss in App.~\ref{app:phasetransition}, for general $q$ this crossing can be identified as a transition in the Dzhaparidze-Nersesyan-Pokrovsky-Talapov (DN-PT) universality class.\cite{dzhaparidze1978magnetic,pokrovsky1979ground,alcaraz1995critical} Fig.~\ref{fig:criticallines} also shows the localisation of the zero modes given by $\ell$ and $r$ in each region, where the first letter corresponds to $T^a_0$ and the second to $T^b_0$.


\subsection{Topological order}
We have discussed the presence of zero modes, the double degeneracy of the ground state, and the spectral gap. We also concluded that the ground states cannot be distinguished by local measurements. These properties do not come by surprise, because of the tight connection to the Kitaev chain. Specifically, at $q=1$ the bulk model \eqref{eq:kitaev} reduces to the Kitaev chain, which will be in its topological phase for all $\eta\neq 1$. It is then easy to see that we can adiabatically change the system away from $q=1$. Starting at $q=1$ for  $\eta>1$, we can make a smooth path to any other $\eta'>1$ and $q>0$ without closing the gap (see Fig.~\ref{fig:criticallines}), thus remaining in the topological phase. The same argument applies to all $\eta<1$.

To support this statement, we consider two topological invariants: the $\mathbb{Z}_2$ invariant\cite{budich2013equivalent,greiter20141d,kawabata2017exact} for class D topological superconductors, and, since we do not explicitly break time reversal symmetry, the $\mathbb{Z}$ invariant\cite{Tewari,Sarkar18} for class BDI.

\emph{$\mathbb{Z}_2$ invariant:} The fermionic parity of the closed system with twisted boundary conditions (TBC) is related to the topological properties of the open system. In general twisted boundary conditions are implemented by adding the boundary Hamiltonian
\begin{align}
h_{\rm bound}=&-\Phi\bigg[t\left(e^{i \varphi_1}c_L\dagg c_{1}+{\rm h.c.}\right)\nonumber\\
&\qquad-\Delta \left(e^{i\varphi_2}c_L\dagg c_{1}\dagg+{\rm h.c.}\right)\bigg]\label{eq:twistedboundary}\\
&-\frac{q_{1}}2(2n_{1}-1)-\frac{q_L}2(2n_L-1).\nonumber
\end{align}
Recall that $t$ and $\Delta$ are given by Eq.~\eqref{eq:tDeltaeta}. The open system corresponds to $\Phi=0$. For $\Phi=1$ we find periodic boundary conditions (PBCs) for $(\varphi_1,\varphi_2)=(0,0)$, and anti-periodic boundary conditions (APBCs) for $(\varphi_1,\varphi_2)=(\pi,\pi)$. In Ref.~\onlinecite{kawabata2017exact} it was shown that a different fermionic parity for the ground states for PBCs and APBCs corresponds to the topological phase, while equal fermionic parity corresponds to the trivial phase. In App.~\ref{app:GSPBCAPBC} we show that the ground states for PBCs and APBCs are  
\begin{equation}
\ket{\Psi^{\rm PBC}}=\ket{\Psi^o},\qquad \ket{\Psi^{\rm APBC}}=\ket{\Psi^e} .
\end{equation}
The parity of $\ket{\Psi^e}$ is $+1$ and the parity of $\ket{\Psi^o}$ is $-1$, which confirms the existence of the topological phase for all $\eta\neq1$.

\emph{$\mathbb{Z}$ invariant:} Using the second invariant we will be able to directly link the topological phase in Fig.~\ref{fig:criticallines} to the topological phase in the conventional Kitaev chain.\cite{kitaev2001unpaired}

In Ref.~\onlinecite{Tewari} it was shown that for a model in class BDI there exists a $\mathbb{Z}$ invariant in the form of a winding number
\begin{equation}
W=-\frac{i}{\pi}\int_{k=0}^{k=\pi}\frac{\mathrm{d} z(k)}{z(k)},
\end{equation}
where $z=\det(A(k))/|\det(A(k))|$. The matrix $A(k)$ is related to the rotated BdG Hamiltonian in $k$-space
\begin{equation}
U\mathcal{H}(k)U\dagg=\begin{pmatrix}
0&A(k)\\
A^\intercal(-k)&0
\end{pmatrix}.
\end{equation}
From the periodic alternating model [i.e. $(\varphi_1,\varphi_2)=(0,0)$ and $\Phi=1$ in Eq.~\eqref{eq:twistedboundary}] we obtain
\begin{equation}
A_q(k)=-\begin{pmatrix}
q&t\cos(k)+i\Delta \sin(k)\\
t\cos(k)+i\Delta \sin(k)&q^{-1}
\end{pmatrix},
\end{equation}
where the $q$-subscript refers to the alternating model. Direct evaluation now yields 
\begin{equation}
W_q=\begin{cases}
1, & t>1,\\
0, & t<1.
\end{cases}
\end{equation}
For the gapped regions ($\eta\neq1$) the system is in the upper case, hence topological. The phase transition $\eta=1$ corresponds to $t=1$, where $W$ is not well defined.

We can also reach this conclusion indirectly, by realising that $\det[A_q(k)]$ does not dependent on $q$. In fact, if one would calculate $A_\textup{Kit}(k)$ for the conventional homogeneous Kitaev chain, but viewed with a two site unit cell, one would find 
\begin{equation}
\det[A_q(k)]=\det[A_\textup{Kit}(k)].
\end{equation}
Consequently, also $W_q=W_\textup{kit}$. This relates the topological phase for general $q$ to the topological phase for the conventional Kitaev chain. Finally we note that the two invariants are related by the fact that the $\mathbb{Z}_2$ invariant is just the parity of the $\mathbb{Z}$ invariant.\cite{Tewari}

\section{Fully inhomogeneous model}\label{sec:fullinhom}
In this section we briefly discuss a model with more general couplings than the alternating setup \eqref{eq:parameters}. Specifically, we consider completely inhomogeneous couplings $\eta_j$ and $q_j$, i.e., the local Hamiltonian takes the form 
\begin{equation}\label{eq:lochamfullinhomgenmaj}
h_j=-\frac{i}2(\eta_j b_j a_{j+1} -\eta_j^{-1} a_jb_{j+1} -q_j a_j b_j-q_j^{-1} a_{j+1}b_{j+1}).
\end{equation}
From the most general ansatz for Majorana zero modes,
\begin{equation}
T^a_0=\sum_{j=1}^L\alpha_ja_j,\qquad T^b_0=\sum_{j=1}^L\beta_jb_j,
\end{equation}
one finds by requiring $[h_j,T^{a,b}_0]=0$ for all $j$ that the coefficients have to satisfy the recursion relations
\begin{equation}
\alpha_{j+1}=-\frac{q_j}{\eta_j}\alpha_j,\qquad \beta_{j+1}=-q_j\eta_j\beta_j.
\end{equation}
The constants $\alpha_1$ and $\beta_1$ are fixed by the normalisation. We note that the localisation of the modes is not clear a priori.

As a trivial but instructive special case one can consider the homogeneous model with $q_j=q$ and $\eta_j=\eta$. This model was originally studied by Hinrichsen and Rittenberg\cite{HINRICHSEN1992350,hinrichsen1992pokrovski,hinrichsen1993quantum} in the context of deformations of XY spin chains. In fact, they continued the work done by Saleur, \cite{saleur1990symmetries} who discussed a spin chain model corresponding to the homogeneous fermionic model introduced in Eq.~\eqref{eq:lochamfullinhomgenmaj} with $\eta=1$.

\begin{figure}[t]
	\centering
		\includegraphics{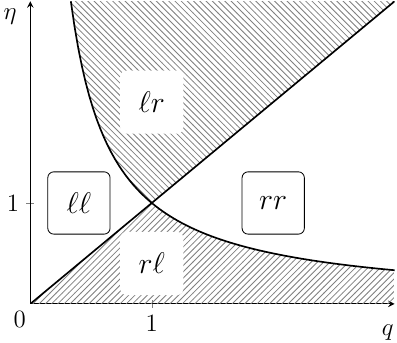}
	\caption{Phase diagram for the homogeneous model \eqref{eq:lochamfullinhomgenmaj} with $\eta_j=\eta$ and $q_j=q$. The shaded region shows the topological phase, characterised by the existence of Majorana zero modes on opposite edges. The notation is as in Fig.~\ref{fig:criticallines}.}\label{fig:phasediagramhomogeneous}
\end{figure}

In the homogeneous setup Hinrichsen and Rittenberg showed that the zero modes simplify to
\begin{eqnarray}
T^a&=&\sqrt{\frac{1-(q/\eta)^2}{1-(q/\eta)^{2L}}}\sum_{j=1}^L\left(-\frac{q}{\eta}\right)^{j-1}a_j,\\ T^b&=&\sqrt{\frac{1-(q\eta)^2}{1-(q \eta)^{2L}}}\sum_{j=1}^L\left(-q\eta\right)^{j-1}b_j.
\end{eqnarray}
Depending on the parameters the modes are localised on the same or opposite edges. The phase diagram can be deduced from the single-particle energies\cite{hinrichsen1993quantum}
\small
\begin{equation}
\Lambda_k=\sqrt{\frac{(q\eta^{-1}-e^{ik})(q\eta^{-1}-e^{-ik})(q\eta-e^{ik})(q\eta-e^{-ik})}{4q^2}}
\end{equation}
\normalsize
for $k=\frac{2\pi n}{L}$ with $n=1,\ldots,L-1$, which imply that the spectral gap closes for $\eta=q$ or $\eta=q^{-1}$. The phase diagram and the localisation of the Majorana zero modes are shown in Fig.~\ref{fig:phasediagramhomogeneous}. 

The gapless lines split the phase space into four regions, each with a different edge mode localisation. To understand which regions are considered topological, we look at the bulk-boundary correspondence such that we can neglect the surface terms. Without the fine tuned surface chemical potential the model reduces to the Kitaev chain, the phase transition in the Kitaev chain directly corresponds to the transitions are $\eta=q$ and $\eta=q^{-1}$. The resulting topological phases are shown as shaded regions in Fig.~\ref{fig:phasediagramhomogeneous}. They are characterised by the appearance of Majorana zero modes at opposite edges. On the other hand, in the trivial phases the zero modes appear on the same edge and are thus not protected against local perturbations. 

We note that the homogeneous model has a richer phase diagram than the model with alternating chemical potential we considered in Sec.~\ref{sec:noninteracting}. However, we stress that the homogeneous model is in general not frustration-free. 

\section{Interacting model}\label{sec:interactions}
In this section we add interactions to the model from Sec.~\ref{sec:noninteracting}. By a special construction we will make the model interacting while keeping the ground states in the disentangled form \eqref{eq:factorizedGS}. Consequently, the interacting model will also prove to be frustration-free. There have been recent developments on frustration-free interacting spinless fermion models.\cite{katsura2015exact,jevtic2017frustration} We will show that in a specific limit we retrieve the model in Ref.~\onlinecite{katsura2015exact}. Moreover, the two-fold degenerate ground state provides us with the notion of a \textsl{weak} zero modes.\cite{alicea2016topological} In the last part of this section we will discuss the phase diagram of the interacting model, giving more insight in the topological order.

\subsection{Hamiltonian}
Recall that the non-interacting system left the fermionic parity invariant, allowing us to split the local Hamiltonian in even- and odd-parity parts [cf. Eqs.~(\ref{eq:hloceven},~\ref{eq:hlocodd})],
\begin{equation}\label{eq:lochamsplit}
h_j=h_j^eP_j^e+h_j^oP_j^o,
\end{equation}
where $P_j^{e,o}$ project onto the even/odd fermion parity sectors of the two-site Hilbert space at lattice sites $j$ and $j+1$. We can define two more projectors, one in each sector, via 
\begin{equation}\label{eq:awayprojectors}
Q_j^{e,o}=\left(h_j^{e,o}+\frac{\mathcal{N}}2\right)P_j^{e,o},
\end{equation}
such that Eq.~\eqref{eq:lochamsplit} becomes
\begin{equation}\label{eq:heodecom3}
h_j=Q_j^e+Q_j^o-\frac{\mathcal{N}}2.
\end{equation} 
The projectors \eqref{eq:awayprojectors} annihilate the ground states \eqref{eq:factorizedGS}. In terms of fermionic operators they are explicitly expressed as
\begin{eqnarray}
Q_j^e&=&-\frac{\eta-\eta^{-1}}2(c_j\dagg c_{j+1}\dagg+c_{j+1}c_j)\nonumber\\
& &-\frac{q_j+q_j^{-1}}2(n_j+n_{j+1}-1)\nonumber\\
& &+\frac{\mathcal{N}}4\bigl[1+(2n_j-1)(2n_{j+1}-1)\bigr],\label{eq:projectoreven}\\
Q_j^o&=&-\frac{\eta+\eta^{-1}}2(c_j\dagg c_{j+1}+c_{j+1}\dagg c_j)\nonumber\\
& &-\frac{q_j-q_j^{-1}}2(n_j-n_{j+1})\nonumber\\
& &+\frac{\mathcal{N}}4\bigl[1-(2n_j-1)(2n_{j+1}-1)\bigr].\label{eq:projectorodd}
\end{eqnarray}
In particular, we see that the projectors \eqref{eq:awayprojectors} both contain a density-density interaction term, which drops out when considering the combination \eqref{eq:heodecom3}.

On the other hand, the existence of the density-density interaction in $Q_j^{e,o}$ points to a way to construct a frustration-free, interacting system. We set
\begin{eqnarray}
H^{\textup{int}}&=&\sum_jh_j^{\textup{int}},\label{eq:Hint}\label{eq:hamint}\\
h_j^{\textup{int}}&=&\sqrt{2}\left[\cos{\phi}\ Q_j^e+\sin{\phi}\ Q_j^o\right]-\mathcal{N}\frac{\cos(\phi-\frac{\pi}4)}2,\nonumber
\end{eqnarray}
where the parameter $\phi$ is restricted to $0<\phi<\pi/2$. The non-interacting model corresponds to the choice $\phi=\pi/4$. We stress that by construction the two factorised states Eq.~\eqref{eq:factorizedGS} are the exact ground states of \eqref{eq:Hint} with energy
\begin{equation}\label{eq:gsenergyinteracting}
E_0=-\frac{(L-1)\cos(\phi-\frac{\pi}4)}2\mathcal{N}.
\end{equation}
Moreover, $H^{\textup{int}}$ is frustration-free, and for $q=1$ it reduces to the model discussed in Ref.~\onlinecite{katsura2015exact}.

Again it is instructive to make the link to spin chains. By applying the Jordan--Wigner transformation \eqref{eq:jordanwigner} we obtain the XYZ chain in an alternating magnetic field
\begin{align}
H^{\textup{int}}=&-\frac{1}{2}\sum_{j=1}^{L-1}\left(J_x\sigma_j^x\sigma_{j+1}^x+J_y\sigma_j^y\sigma_{j+1}^y +J_z\sigma_j^z\sigma_{j+1}^z\right)\nonumber\\
 &\qquad -\sum_{j=1}^LB_j\sigma_j^z\label{eq:XYZ},
\end{align}
where 
\begin{equation}\label{eq:JxJyJz}
J_x=\rho\eta+\varrho\eta^{-1},\quad J_y=\rho\eta^{-1}+\varrho\eta,\quad J_z=\varrho\mathcal{N},
\end{equation}
with $\rho=\cos(\phi-\pi/4)$ and $\varrho=\sin(\phi-\pi/4)$, and 
\begin{equation}\label{eq:B0B1}
B_j=\begin{cases}
B_0=\rho q-\varrho q^{-1},&\textup{ if $j$ odd,}\\
B_1=\rho q^{-1}-\varrho q,&\textup{ if $j$ even.}
\end{cases}
\end{equation}
It turns out that these parameters satisfy the following condition
\begin{equation}\label{eq:equality}
B_0B_1=J_z^2+J_xJ_y-J_z\sqrt{(J_x+J_y)^2+(B_0-B_1)^2}.
\end{equation}
This shows great resemblance to the frustration-free condition for homogeneous XYZ chains provided by Refs.~\onlinecite{muller1985implications}~and~\onlinecite{KURMANN1982235}, in which case the condition becomes $B^2=(J_x-J_z)(J_y-J_z)$. Thus it seems that the frustration-free condition in the alternating case is indeed given by Eq.~\eqref{eq:equality}.

\begin{figure}[t]
		\includegraphics{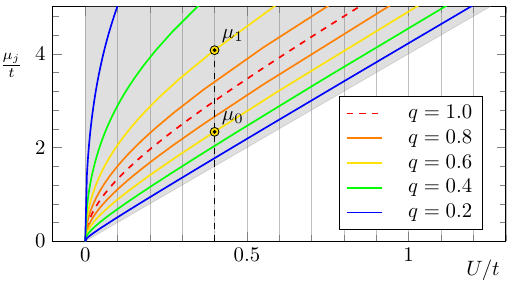}
	\caption{(Colour online) Frustration-free lines in $U-\mu$ space for $t=\Delta$. The Peschel--Emery line\cite{katsura2015exact,Peschel1981}  for the homogeneous case is given by the red dashed line. The other lines depict the alternating model. For every $q$ the two chemical potentials in the system are given as a function of the interaction $U$. As an example $q=0.6$ and $U=0.4$ is explicitly shown, the two chemical potentials are indicated by the two yellow dots. The vertical lines are guides to the eye, each $U$ relates to two $\mu_j$'s.}\label{fig:Peschelemery}
\end{figure}
It is illustrative to return to the language of the interacting Kitaev chain 
\begin{align}
H=&-\sum_{j}\left[t(c\dagg_jc_{j+1}+c\dagg_{j+1}c_j)+\Delta(c\dagg_jc_{j+1}^\dagger+c_{j+1}c_j)\right]\nonumber\\
&-\frac{1}{2}\sum_j\mu_j(2 n_j-1)+U\sum_j(2n_j-1)(2n_{j+1}-1),
\end{align}
where the parameters $t$, $\Delta$, $\mu_j$ and $U$ are non-trivial functions of $q$, $\eta$ and $\phi$. The chemical potential alternates between the values 
\begin{equation}
\mu_j=\begin{cases}
\mu_0=\mu_0(q,\eta,\phi),&\textup{ if $j$ odd,}\\
\mu_1=\mu_1(q,\eta,\phi),&\textup{ if $j$ even.}
\end{cases}
\end{equation}
The condition of the model to be frustration-free results in relations between the parameters $t$, $\Delta$, $\mu_{0,1}$ and $U$. For example, in Fig.~\ref{fig:Peschelemery} we set $t=\Delta$ which fixes the function $\eta(\phi)$, which in turn determines the chemical potential and interaction as functions of $q$ and $\phi$, i.e., $\mu_{0,1}(q,\phi)$ and $U(q,\phi)$. Inverting the latter relation we obtain the conditions on the chemical potentials $\mu_{0,1}(q,U)$ for the model to become frustration-free, which is plotted in Fig.~\ref{fig:Peschelemery} for different values of $q$. One way to view this result is that given an interaction strength $U$ and the chemical potential $\mu_0$ on the odd sites we can determine the  inhomogeneity parameter $q$ and thus the chemical potential $\mu_1$ on the even sites. In the homogeneous case $q=1$ we recover the Peschel--Emery line.\cite{Peschel1981,katsura2015exact} In the limit $q\rightarrow0$ one of the chemical potentials diverges, and the other approaches $4U$.

\subsection{Weak zero modes}
In general, if the ground state is degenerate one can always find operators mapping one ground state to the other. Some interacting systems allow for strong zero modes, commuting with the full Hamiltonian.\cite{kells2015multiparticle,alexandradinata2016parafermionic,fendley2016strong} However, in general interactions destroy this feature and only the commutation within the low-energy sector remains. For the system \eqref{eq:Hint} we already know the exact form of the zero modes, because for $\ket{\Psi^{e,o}}$ we showed that $T^{a,b}_0\ket{\Psi^{e}}=\ket{\Psi^{o}}$ (see App.~\ref{app:actionzeromodes}). However, in the interacting model we have $[H^\text{int},T^{a,b}_0]\neq 0$, thus the modes $T^{a,b}_0$ are weak zero modes as defined in Ref.~\onlinecite{alicea2016topological}. 

\subsection{Phase Diagram}
In this section we discuss the phase diagram of the interacting model \eqref{eq:Hint}. Without interactions it was possible to obtain exact results for the ground-state energy density and the spectral gap. When adding interactions generically one loses the analytical expressions for the observables and the only rescue lies in numerical tools. However, due to the specific construction for the interacting model, the ground-state energy can still be found analytically as we saw in Eq.~\eqref{eq:gsenergyinteracting}. For the spectral gap there is no analytic solution. Nevertheless, we can find lower and upper bounds for the gap by using the min-max principle.\cite{horn1990matrix} This will give an indication for the gapped and gapless regions in phase space. To confirm these results, and fill in the remaining blanks we also perform a numerical analysis.

We start by discussing the bounds on the spectral gap. Recall that the local Hamiltonians are given by
\begin{equation}
h_j^{\textup{int}}=\sqrt{2}\left[\cos{\phi}\ Q_j^e+\sin{\phi}\ Q_j^o\right].
\end{equation}
We are not concerned with the constant term, because we are interested in the energy gap. Since $Q^{e,o}$ are projection operators, they are positive semidefinite. We introduce the notion of operator inequality as: $A\geq B$ if $A-B$ is positive semidefinite. Two cases have to be distinguished: (i) $0<\phi<\pi/4$ and (ii) $\pi/4<\phi<\pi/2$.  In case (i), we have
\begin{equation}
\sqrt{2} \sin \phi\ (Q^e_j + Q^o_j) \le h^{\rm int}_j 
\le \sqrt{2} \cos \phi\ (Q^e_j + Q^o_j).
\end{equation}
From Eq.~\eqref{eq:lochamsplit} we recognise that $Q^e_j + Q^o_j$ is nothing but a local Hamiltonian of the non-interacting model (up to a constant shift). This allows us to write
\begin{equation}
\sqrt{2} \sin \phi\, H \le H^{\rm int} \le \sqrt{2} \cos \phi\, H,
\end{equation}
where $H^{\rm int}=\sum^{L-1}_{j=1} h^{\rm int}_j$.
Then, the min-max principle tells us that\cite{horn1990matrix}
\begin{equation}
\sqrt{2} \sin \phi\ E_n \le E^{\rm int}_n \le \sqrt{2} \cos \phi\ E_n,
\end{equation}
where $E_n$ and $E^{\rm int}_n$ ($n=1,2,3,\ldots$) are $n$th eigenvalue of $H$ and $H^{\rm int}$, respectively. Since the interacting and the non-interacting Hamiltonians share the same ground states annihilated by all $Q^{e,o}_j$, we have
\begin{equation}\label{eq:bound1}
\sqrt{2} \sin \phi \, \Delta E \le \Delta E^{\rm int} \le \sqrt{2} \cos \phi \, \Delta E,
\end{equation}
where $\Delta E=E_3$ is the energy gap of the non-interacting system introduced in Eq.~\eqref{eq:noningap}, while $\Delta E^{\rm int}=E^{\rm int}_3$ is the one of the interacting system. Repeating the same argument, we find that the gap in case (ii) is bounded as
\begin{equation}\label{eq:bound2}
\sqrt{2} \cos \phi \, \Delta E \le \Delta E^{\rm int} \le \sqrt{2} \sin \phi \, \Delta E.
\end{equation}
Concluding, by using the min-max principle we have found upper and lower bounds on the gap energy for the interacting system.

From these bounds we can already draw several conclusions. First of all for $\eta\neq1$ and $\phi\neq 0,\pi/2$ the system is gapped. The lower bound is finite, since both $\sin \phi/\cos \phi$ and $\Delta E$ are positive. Also, for $\eta=1$ the gap has to close, independent of the interaction (governed by $\phi$). Approaching this point both the upper and lower bound vanish, since both are proportional to $\Delta E$ (the non-interacting gap). In the following part we will discuss numerical results to confirm these statements. Moreover, there are two boundaries ($\phi=0,\pi/2$), that cannot be addressed by the above reasoning. At these points the lower bound vanishes, while the upper bound is finite. We will come back to these special points below.

\begin{figure}[t]
	\includegraphics{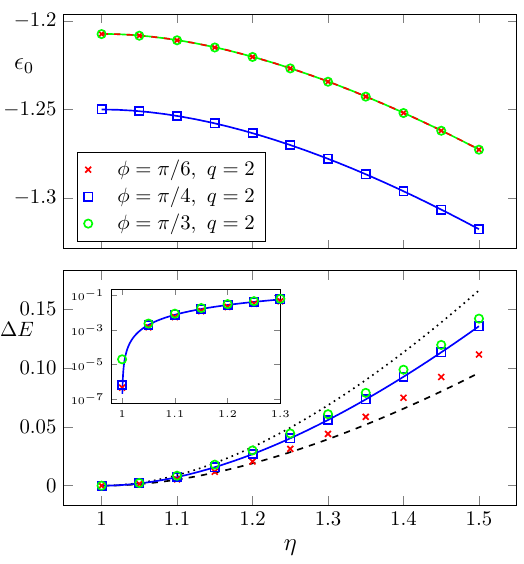}
	\caption{(Colour online) Ground-state energy per lattice site, $\epsilon_0=E_0/L$, (top panel) and spectral gap (bottom panel) as a function of $\eta$ obtained numerically (DMRG, bond dimension $D=16$, results for system sizes $L=40,80,\ldots,160$ extrapolated to the thermodynamic limit) for $q=2$ and for three values for the interaction parameter $\phi$. The non-interacting $\phi=\pi/4$ results are depicted by blue squares, the two interacting cases $\phi=\pi/6,\pi/3$ by red crosses and green circles respectively. In the top panel, the analytical results for ground-state energy are shown as solid lines [see Eq.~\eqref{eq:gsenergyinteracting}, red and green are overlapping]. In the bottom panel, the analytical spectral gap for $\phi=\pi/4$ is shown as the blue line [Eq.~\eqref{eq:noningap}]. The inset shows the same results on a log scale, to emphasize the gap closing at $\eta=1$. The lower and upper band are depicted by respectively the dashed and dotted line [Eqs.~(\ref{eq:bound1},~\ref{eq:bound2})].}\label{fig:DMRGeta}
\end{figure}

\begin{figure}[t]
\includegraphics{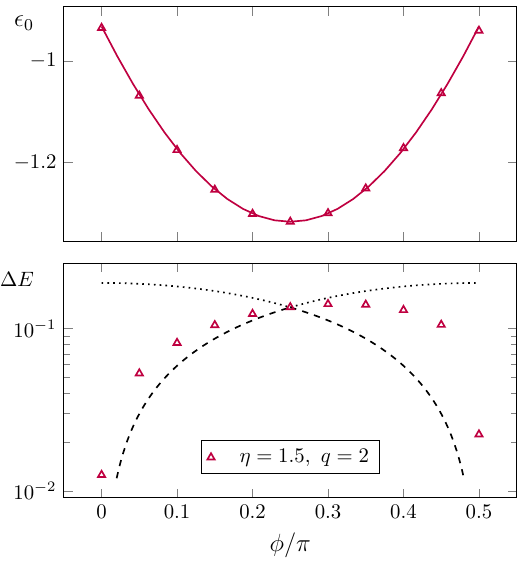}
	\caption{(Colour online) Ground-state energy per lattice site, $\epsilon_0=E_0/L$, (top panel) and spectral gap (bottom panel) as a function of $\phi$ obtained numerically for $q=2$ and $\eta=1.5$. The purple triangles depict the DMRG results, for the $\epsilon_0$ also the analytical results are shown by the solid line [Eq.~\eqref{eq:gsenergyinteracting}]. The lower and upper band are depicted by respectively the dashed and dotted line [Eqs.~(\ref{eq:bound1},~\ref{eq:bound2})]. The DMRG calculations were performed with the parameters of Fig.~\ref{fig:DMRGeta}. }\label{fig:DMRGphi}
\end{figure}

We use a density-matrix renormalisation group (DMRG) algorithm to explore the low-energy spectrum in parameter space.\cite{white1992density,white1993density,SCHOLLWOCK201196} Using finite-size scaling we obtained ground-state energy and spectral gap in the thermodynamic limit. Examples of these results (for $q=2$) are shown in Figs.~\ref{fig:DMRGeta} and~\ref{fig:DMRGphi}. The top figures show that the exact and numerical findings for the ground-state energy match perfectly. 

The bottom panel of Fig.~\ref{fig:DMRGeta} shows the numerical results for the energy gap $\Delta E$ between the ground states and the first excited state. For the non-interacting case ($\phi=\pi/4$) also the analytic result is depicted by the solid blue line. For the interacting cases the dashed (dotted) line depicts the lower (upper) bound on the gap energy, confirming that the gap lies between the two bounds. As expected, for all interaction parameters $\phi$ the system only becomes gapless at $\eta=1$. This we can clearly see in the inset, where the gap is depicted on a logarithmic scale.\footnote{Here we have to note that the predicted gap at $\eta=1$ vanishes, however the DMRG can never truly reach zero, due to the diverging entanglement at critical points. This is the only numerical point lying outside the bounds.}  Also, the lower panel of Fig.~\ref{fig:DMRGphi} confirms that, away from $\eta=1$, the gap does not close for $0<\phi<\pi/2$. From Fig.~\ref{fig:DMRGphi} we can also deduce what happens when approaching  the extremal cases $\phi=0$ and $\phi=\pi/2$. The bounds do not converge (to zero), nevertheless, the gap closes when approaching either boundary. 

\begin{figure}[t]
	\centering
	\includegraphics{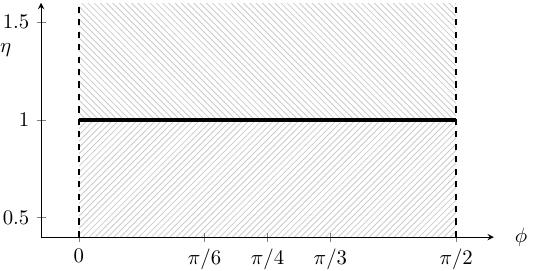}
	\caption{Phase diagram of the interacting model \eqref{eq:Hint} in $\eta-\phi$ plane. The solid black line denotes a phase transition in the DN-PT universality class, while at the dashed black lines the ground state becomes highly degenerate due to additional symmetry (see main text). The phase diagram is independent of the inhomogeneity parameter $q$.} \label{fig:phasediagrametaphi}
\end{figure}
The phase diagram of the interacting model is shown in Fig.~\ref{fig:phasediagrametaphi}. The solid line represents the phase transition, which, as we argue in App.~\ref{app:phasetransition}, is in the DN-PT universality class (like the non-interacting model).\cite{alcaraz1995critical} At the dashed lines the ground state becomes highly degenerate due to additional symmetry. We stress that the phase diagram is independent on the inhomogeneity parameter $q$. 

Finally, let us discuss what happens for $\phi=0,\pi/2$. In the phase diagram Fig.~\ref{fig:phasediagrametaphi} they correspond to the dashed black lines. For these specific parameters only the even or odd two-site projector [Eq.~\eqref{eq:awayprojectors}] is present in the Hamiltonian. As we will see, this induces an additional symmetry in the low-energy sector, which in turn raises the ground-state degeneracy to $L+1$. Here we explicitly show the construction of the ground states for $\phi=\pi/2$, the derivation for $\phi=0$ is similar. We consider two different cases $q=1$ and $q\neq1$. 

\emph{Case 1: $q=1$}: The local Hamiltonian reduces to the isotropic Heisenberg term if we rewrite the model in spin language using Eq.~\eqref{eq:jordanwigner},
\begin{equation}
h_j^\text{int}=-\frac{\eta+\eta^{-1}}{2\sqrt{2}}\left[\sigma^x_j\sigma^x_{j+1}+\sigma^y_j\sigma^y_{j+1}+\sigma^z_j\sigma^z_{j+1}\right].
\end{equation}
The Hamiltonian possesses an sl$_2$-symmetry generated by
\begin{equation}
\mathcal{S}^z=\sum_j\sigma_j^z,\qquad\mathcal{S}^\pm=\sum_k\sigma_j^\pm,\label{eq:Szpm}
\end{equation}
where $\sigma^\pm=\frac{\sigma^x\pm i \sigma^y}2$. If we define the fully polarised state $\ket{\Uparrow}=\ket{\su\su\ldots\su}$, then the ground states are given by
\begin{equation}
\ket{\Psi_{p}}=(\mathcal{S}^-)^p\ket{\Uparrow},\quad p=0,1,\ldots,L,
\end{equation}
which are the $L+1$ states with $S_{\rm tot}=L/2$.

\emph{Case 2: $q\neq1$}: The model can be represented as an XXZ chain with alternating magnetic field,
\begin{align}
h_j^\text{int}=&-\frac{\mathcal{N}}{2\sqrt{2}}\Bigg[\frac{\eta+\eta^{-1}}{\mathcal{N}}\left(\sigma_j^x\sigma_{j+1}^x+\sigma_{j}^y\sigma_{j+1}^y\right)+\sigma_{j}^z\sigma_{j+1}^z\nonumber\\
& +\frac{q_j-q_j^{-1}}{\mathcal{N}}\left(\sigma_j^z-\sigma_{j+1}^z\right)-1\Bigg],\label{eq:lochamint2}
\end{align}
where we have shifted the spectrum such that we have a zero-energy ground state. We note in passing that the model shows great resemblance to the XXZ model studied in Ref.~\onlinecite{PASQUIER1990523}, which supports U$_q($sl$_2)$ symmetry. In that case the ground state is also $L+1$-fold degenerate. However, because of the quantum group symmetry, the excited states also have additional degeneracies which are not observed in the spectrum of Eq.~\eqref{eq:lochamint2}. Nevertheless, we can still construct the lowering operator analogous to $\mathcal{S}^-$ in Eq.~\eqref{eq:Szpm} to derive the $L+1$ ground states.

Note that the polarised state $\ket{\Uparrow}$ is still a ground state. We can define a lowering operator
\begin{equation}\label{eq:lowering}
\tilde{\mathcal{S}}^-=\sum_j g_j\sigma_j^-,\qquad 
g_j=\begin{cases}
\sqrt{1-\frac{q-q^{-1}}{\mathcal{N}}}, \quad j \textup{ odd,}\\
\sqrt{1+\frac{q-q^{-1}}{\mathcal{N}}}, \quad j \textup{ even.}
\end{cases}
\end{equation}
The coefficients $g_j$ have been obtained recursively ensuring that $h_j\tilde{\mathcal{S}}^-\ket{\Uparrow}=0$. Since $(\sigma^-_j)^2=0$ we see that $(\tilde{\mathcal{S}}^-)^{L+1}=0$. Furthermore, it turns out that 
\begin{equation}\label{eq:case2groundstates}
\ket{\Psi_p}=(\tilde{\mathcal{S}}^-)^p\ket{\Uparrow}
\end{equation}
for $p=0,1,\ldots,L$ are ground states. The ground states $(\tilde{\mathcal{S}}^-)^p\ket{\Uparrow}$ are non-vanishing. This is because, away from the phase transition ($\eta=1$), the coefficients $g_j$ are strictly positive, hence there is no destructive interference for $p\leq L$, when acting on a single state ($\ket{\Uparrow}$).

We can check that the Hamiltonian still leaves 
\begin{equation}\label{eq:Sz}
S^z=\mathcal{S}^z/2
\end{equation}
invariant. Since $\ket{\Psi_p}$ belongs to the $S^z=L/2-p$
 sector, the ground states are linearly independent. Details of the construction of this operator and the proof of Eq.~\eqref{eq:case2groundstates} are given in App.~\ref{app:oddprojector}.

\section{Conclusion}
In this article we have investigated frustration-free topological systems. Specifically, we studied non-interacting and interacting generalisations of the Kitaev chain, with alternating chemical potential on the lattice sites. Both introduced models possess two exactly degenerate ground states of product form. This allowed us to determine exactly the Majorana zero modes mapping the ground states onto each other. Only in the non-interacting case, these modes commute with the full Hamiltonian, making them \emph{strong} zero modes. We stress that due to a fine-tuned boundary term all our results are exact even for finite systems, which is in contrast to the generic Kitaev chain where the Majorana edge mode energy only vanishes exponentially with the system length. For the non-interacting model we have shown that there is a finite energy gap above the ground states, except at the phase transition ($\eta=1$) given by the zero-pairing limit. Hence there exists a smooth path connecting the inhomogeneous model to the (homogeneous) Kitaev chain, proving that both are in the same topological phase. Also, we have shown both analytically and numerically that the interacting model remains gapped in a certain region, implying that the interacting model is in the same topological phase as the corresponding non-interacting model.

In the future it would be interesting to investigate whether frustration-free models can also be constructed for genuinely interacting systems like $\mathbb{Z}_n$ clock models.\cite{iemini2017topological,xu2017matrix,mazza2013robustness,mazza2018poor,mahyaeh2018exact} It would also be interesting to generalise our interacting model to include non-hermiticity. In a recent work, it was shown that a non-hermitian one-dimensional
spinless p-wave superconductor can support complex edge modes in addition to Majorana zero modes.\cite{kawabata2018parity} These give rise to a purely imaginary shift in energy. Future work could be dedicated to discussing a non-hermitian extension of the models discussed in this article.

\section*{Acknowledgment}
We would like to thank Eddy Ardonne, Kohei Kawabata and Iman Mahyaeh for useful comments on the manuscript. 
This work was supported by the Foundation for Fundamental Research on Matter (FOM), which is part of the Netherlands Organisation for Scientific Research (NWO), under 14PR3168. H.K. was supported in part by JSPS KAKENHI Grant Nos. JP18H04478, JP18K03445. D.S. acknowledges support of the D-ITP consortium, a program of the Netherlands Organisation for Scientific Research (NWO) that is funded by the Dutch Ministry of Education, Culture, and Science (OCW).

\appendix
\section{Lindblad operators}\label{app:lindblad}
An alternative approach in finding the ground states of the non-interacting Hamiltonian is by virtue of Lindblad operators. Less steps are required for deriving that $\ket{\Psi^\pm}$ are the ground states of $H$, it is, however, a less transparent method. A similar approach has been used by Tanaka,\cite{tanaka2016one} who discussed a more general model with interaction, which includes as a special case the non-interacting model.

Lindblad operators $L_j$ are defined in the context of non-equilibrium dynamics and govern the dissipation in the system.\cite{schaller2014open,Diehl2011,kraus2008preparation,johnson2016general} For certain states $\ket{D}$, called dark states in the quantum optics literature, this dissipation term vanishes, which is expressed in terms of the Lindblad operators as 
\begin{equation}
L_j\ket{D}=0.
\end{equation}
Here we leave the dissipation picture and use the notion of dark states to define
\begin{equation}
L_j=\sqrt{g}\left(x_j^{-1} c_j-x_j c_j\dagg-x_{j+1}^{-1} c_{j+1}-x_{j+1} c_{j+1}\dagg\right),
\end{equation}
where $g=\frac{\eta^2-\eta^{-2}}{\mathcal{N}}$ and $x_j=x_0$ ($x_1$) for $j$ odd (even). One can now easily see that the dark states of $L_j$ are given by $\ket{\Psi^\pm}$ as defined in Eq.~\eqref{eq:factorizedGS}. Furthermore, the Lindblad operators are chosen such that
\begin{equation}\label{eq:applindbladham}
H=\sum_j h_j,\qquad h_j=L_j\dagg L_j -\frac{\mathcal{N}}2,
\end{equation}
with $h_j$ the non-interacting Hamiltonian in Eq.~\eqref{eq:lochamfermion}.
Hence we have found two ground states of the Hamiltonian. In the following we will drop the overall constant $\mathcal{N}/2$.

Assuming that we have no knowledge of the ground state degeneracy from zero modes or the like, we still have to show that we have found all ground states. This can be done by Witten's conjugation argument:\cite{hagendorf2013spin,witten1982constraints} The model we are interested in has at least a two-fold degenerate ground state and is given by
\begin{equation}
H=\sum_jL_j\dagg L_j.
\end{equation}
Now consider an invertible matrix $\mathcal{M}$, then we can define $\tilde{L}_j=\mathcal{M}L_j\mathcal{M}^{-1}$ such that 
\begin{equation}
\tilde{H}=\sum_j\tilde{L}_j\dagg \tilde{L}_j
\end{equation}
has the same number of ground states as $H$. If we choose
\begin{equation}
\mathcal{M}=[1+(x_0^{-1}-1)n_1][1+(x_1^{-1}-1)n_2]\ldots[1+(x_1^{-1}-1)n_L]
\end{equation}
then $\tilde{L}_j=\sqrt{g}\left(c_j-c_j\dagg- c_{j+1}- c_{j+1}\dagg\right)$ and the conjugated Hamiltonian becomes
\begin{equation}
\tilde{H}=-2g\sum_j\left[c_j\dagg c_{j+1}+c_{j+1}\dagg c_j+ c_{j+1}c_j+c_j\dagg c_{j+1}\dagg \right],
\end{equation}
which is nothing but the Kitaev chain in the limit $t=\Delta$ and $\mu=0$, which clearly has two ground states.

\section{Local operator and correlation length}\label{app:correlators}
In order to determine the correlation length, we calculate the equal-time Green function $G^{e,o}(i,j)=\braket{\Psi^{e,o}|c_i\dagg c_j|\Psi^{e,o}}$. Note that $G^{e,o}$ does depend on the specific $i$ and $j$ and not only on the distance, because translational invariance is broken. If we define $d=|i-j|$, the Green function can be written as
\begin{equation}
G^{e,o}(i,j)=\frac{x_ix_j}{(1-x_i^4)(1-x_j^4)}\left[\frac{\eta^{-d}}{1\pm\eta^L}+\frac{\eta^{d}}{1\pm\eta^{-L}}\right],
\end{equation}
where the upper sign corresponds to $G^e(i,j)$ and the lower to $G^{o}$. Furthermore, $x_i=x_0$ for $i$ odd and $x_i=x_1$ for $i$ even, with $x_{0,1}$ defined in Eqs.~(\ref{eq:x0},~\ref{eq:x1}), and we have used the identity
\begin{equation}\label{eq:x0x1identity1}
\frac{x_0^2+1}{x_0^2-1}\frac{x_1^2+1}{x_1^2-1}=\eta^2.
\end{equation}
For large $L$ the Green function is proportional to
\begin{equation}
G^{e,o}(i,j)\propto\begin{cases}
\eta^d &\eta<1,\\
\eta^{-d}&\eta>1,
\end{cases}
\end{equation}
which scales as $e^{-d/\xi}$ with the correlation length
\begin{equation}\label{eq:correlationlength}
\xi=\frac1{\log[\max(\eta,\eta^{-1})]}.
\end{equation}
We note that $\xi$ diverges at the phase transition ($\eta=1$).

Next we show that the difference between the expectation values of an even local operator $O_e$ with respect to $\ket{\Psi^e}$ and $\ket{\Psi^o}$ satisfies the bound \eqref{eq:difference}. First we recognise that for an even local operator $O_e$
\begin{equation}
\bigl[O_e,(-1)^F\bigr]=0,
\end{equation}
and recall that $\ket{\Psi^-}=(-1)^F\ket{\Psi^+}$. Therefore we can already infer that
\begin{align}
\braket{\Psi^-|O_e|\Psi^-}&=\braket{\Psi^+|O_e|\Psi^+},\\
\braket{\Psi^-|O_e|\Psi^+}&=\braket{\Psi^+|O_e|\Psi^-}.
\end{align}
This simplifies the left-hand side of Eq.~\eqref{eq:difference} to 
\begin{align}
&\frac{|N\braket{\Psi^+|O_e|\Psi^-}-M\braket{\Psi^+|O_e|\Psi^+}|}{|N^2-M^2|}\nonumber\\
&\leq\frac{N|\braket{\Psi^+|O_e|\Psi^-}|+M|\braket{\Psi^+|O_e|\Psi^+}|}{|N^2-M^2|}\nonumber\\
&\leq\frac{\norm{O_e}NM+N|\braket{\Psi^+|O_e|\Psi^-}|}{|N^2-M^2|},\label{eq:estimation}
\end{align}
where $N$ and $M$ are given by Eqs.~(\ref{eq:N},~\ref{eq:M}) respectively. Here $\norm{O_e}$ is the operator norm defined as
\begin{equation}\label{eq:operatornorm}
\norm{O_e}:=\inf\{c\ : \ \norm{A\ket{\phi}}\leq c\norm{\ket{\phi}} \textup{ for all } \ket{\phi}\in\mathbb{C}^{\otimes L}\}.
\end{equation}
In the last line we have resolved one of the two correlators. The other one is more involved. In the following we will derive an upper bound for $|\braket{\Psi^+|O_e|\Psi^-}|$, which becomes a bit intricate because of the inhomogeneous nature of the system. The estimation hinges on the fact that $O_e$ is a local operator with a support on $\ell$ sites.

First of all, notice that there are $L-\ell$ sites for which $\ket{\Psi^\pm}$ commutes with $O_e$, therefore we can reduce 
\begin{align}
\braket{\Psi^+|O_e|\Psi^-}=&C_1\left[(x_0^2-1)(x_1^2-1)\right]^{\frac{L}{2}-\lfloor\frac{\ell}2+1\rfloor}\nonumber\\
&\quad\times\braket{\tilde{\Psi}^+|O_e|\tilde{\Psi}^-},
\end{align}
where 
\begin{equation}
C_1=\max(x_0^2+1,x_0^2-1)\max(x_1^2+1,x_1^2-1),
\end{equation}
and $\lfloor\cdot\rfloor$ is the floor function. Both the constant and the floor are a result of the alternating pattern in the model. They only contribute marginally to magnitude, but for completeness we will keep them in.
Furthermore, $\tilde{\Psi}^\pm$ are the ground states reduced to $i\in[j_1,j_k]$.

Using the Schwarz's inequality we obtain
\begin{align}
|\braket{\tilde{\Psi}^+|O_e|\tilde{\Psi}^-}|^2&\leq\braket{\tilde{\Psi}^+|\tilde{\Psi}^+}\braket{\tilde{\Psi}^-|O_e\dagg O_e|\tilde{\Psi}^-}\\
&\leq \left(C_2\left[(x_0^2+1)(x_1^2+1)\right]^{\lfloor\frac{\ell}2\rfloor}\norm{O_e}\right)^2,\nonumber
\end{align}
with 
\begin{equation}
C_2=\max(x_0^2+1,1)\max(x_1^2+1,1).
\end{equation}
Plugging this all in yields
\begin{align}
N|\braket{\Psi^+|O_e|\Psi^-}|&\leq NM C_3 \left(\frac{(x_0^2+1)(x_1^2+1)}{(x_0^2-1)(x_1^2-1)}\right)^{\lfloor\ell/2\rfloor}\norm{O_e}\nonumber\\
&=NM C_3 \eta^{2\lfloor\ell/2\rfloor}\norm{O_e},
\end{align}
using Eq.~\ref{eq:x0x1identity1} and defining $C_3=\frac{C_1C_2}{(x_0^2-1)(x_1^2-1)}$. Hence we find
\begin{align}
|\braket{\Psi^e|O_e|\Psi^e}-\braket{\Psi^o|O_e|\Psi^o}|&\leq\frac{(1+C_3 \eta^{2\lfloor\ell/2\rfloor})\norm{O_e}}{|\eta^L-\eta^{-L}|}\label{eq:boundfinal}\\
&\approx(1+C_3\eta^{2\lfloor\ell/2\rfloor})\norm{O_e}e^{-L/\xi},\nonumber
\end{align}
where we have used $N/M=\eta^L$ and $\xi$ is defined in Eq.~\eqref{eq:correlationlength}. For $\eta\neq1$ and $l\ll L$ (i.e., $O_e$ local) the numerator is small compared to $e^{L/\xi}$. Hence Eq.~\eqref{eq:boundfinal} vanishes for large systems.

\section{Action of zero modes}\label{app:actionzeromodes}
In this appendix we will explicitly derive that $T_0^a$ (analogously one can show this for $T_0^b$) maps one ground state to the other, i.e., $T_0^a\ket{\Psi^e}\propto\ket{\Psi^o}$.

Before we proceed with the derivation let us note that
\begin{equation}\label{eq:x0x1identity2}
\frac{x_1(x_0^2-1)}{x_1^2+1}=x_0q\eta^{-1},
\end{equation}
which we will need later on.

Suppose $\ket{\Psi^e}=\sum_j\lambda_ja_j\ket{\Psi^o}$, then $\braket{\Psi^o|a_j|\Psi^e}=\lambda_j$, because $a_j^2=1$. Here we have assumed $\braket{\Psi^o|a_ja_k|\Psi^o}=0$ for $j\neq k$, which is true because $\braket{\Psi^\pm|a_ja_k|\Psi^\pm}=\braket{\Psi^\pm|a_ja_k|\Psi^\mp}=0$ due to the specific construction of the ground states.

And we define the shorthand notation $\ket{(\pm)_j}=(x_j c_j\dagg\pm1)\vac$ (with $x_j=x_{0,1}$ depending on parity), such that for instance $\ket{\Psi^+}=\ket{(++\ldots+)}$. Using this notion 
\begin{align}
c_j\ket{\Psi^\pm}&=(-1)^{j-1}\ket{(\mp\mp\ldots\mp)}(x_j)\vac_j\ket{(\pm\pm\ldots\pm)},\\
c_j\dagg\ket{\Psi^\pm}&=(-1)^{j-1}\ket{(\mp\mp\ldots\mp)}(\pm c_j\dagg)\vac_j\ket{(\pm\pm\ldots\pm)},
\end{align}
which in turn yields
\begin{align}
\braket{\Psi^\pm|c_j|\Psi^\pm}&=\braket{\Psi^\pm|c_j\dagg|\Psi^\pm}\\
&=\pm  x_j(-1)^{j-1}\prod_{k=1}^{j-1}(x_k^2-1)\prod_{l=j+1}^{L}(x_l^2+1),\nonumber\\
\braket{\Psi^\mp|c_j|\Psi^\pm}&=-\braket{\Psi^\mp|c_j\dagg|\Psi^\pm}\\
&=\mp x_j(-1)^{j-1}\prod_{k=1}^{j-1}(x_k^2+1)\prod_{l=j+1}^{L}(x_l^2-1).\nonumber
\end{align}
Recall that $a_j=c_j+c_j\dagg$ (from Eq.~\eqref{eq:majorana}) and $\ket{\Psi^{e,o}}=\frac1{\sqrt{2(N\pm M)}}\left(\ket{\Psi^+}\pm\ket{\Psi^-}\right)$ (from Eqs.~(\ref{eq:GSnorme},~\ref{eq:GSnormo})) such that
\begin{align}
\bra{\Psi^o}a_j\ket{\Psi^e}&=\frac{2x_j(-1)^{j-1}}{\sqrt{N^2-M^2}}\prod_{k=1}^{j-1}(x_k^2-1)\prod_{l=j+1}^{L}(x_l^2+1)\nonumber\\
&=\frac{2x_j(-1)^{j-1}}{\sqrt{N^2-M^2}}\frac{N}{x_j^2+1}\prod_{k=1}^{j-1}\frac{x_k^2-1}{x_k^2+1}.
\end{align}
If $j$ is odd then
\begin{align}
\lambda_j&=\frac{2}{\sqrt{1-\eta^{-2L}}}\frac{x_0}{x_0^2+1}\left(-\sqrt{\frac{(x_0^2-1)(x_1^2-1)}{(x_0^2+1)(x_1^2+1)}}\right)^{j-1}\nonumber\\
&=\frac{2}{\sqrt{1-\eta^{-2L}}}\frac{x_0}{x_0^2+1}\left(-\eta^{-1}\right)^{j-1},
\end{align}
using Eq.~\eqref{eq:x0x1identity1}. For even $j$ we get
\begin{align}
\lambda_j&=-\frac{2}{\sqrt{1-\eta^{-2L}}}\frac{1}{x_0^2+1}\frac{x_1(x_0^2-1)}{x_1^2+1}\left(-\eta^{-1}\right)^{j-2}\nonumber\\
&=\frac{2}{\sqrt{1-\eta^{-2L}}}\frac{x_0}{x_0^2+1}\left[-q\eta^{-1}\left(-\eta^{-1}\right)^{j-2}\right],
\end{align}
using Eq.~\eqref{eq:x0x1identity2}. Comparing these results to Eq.~\eqref{eq:ZMa} we conclude that $\lambda_j=K\alpha_j$, for some constant $K$.  Finally, we check that the $\lambda_j$'s are normalised correctly:
\begin{equation}
\braket{\Psi^e|\Psi^e}=\sum_j \lambda_j^2=\frac{4x_0^2}{(x_0^2+1)^2}\frac{\eta^2+q^2}{\eta^2-\eta^-2}.
\end{equation}
By simply plugging in $x_0$ one can verify that this equals unity. Hence $T_0^a$ maps one ground state to the other, up to an overall phase. For $T_0^b$ one can do a similar derivation.

\section{Spectrum}\label{app:spectrumnonint}
In this appendix we derive the spectrum for the non-interacting model. The first eigenvalue we have already encountered, since there is a zero-energy mode in the system. The other eigenvalues we find by diagonalising $C$ in Eq.~\eqref{eq:C}. We use the following ansatz for the eigenstates of $C$:
\begin{align}
T^a_k(2l-1)&=
\alpha e^{i k (2l-1)}+\bar{\alpha}e^{-ik(2l-1)},\\
T^a_k(2l)&=
\beta e^{i k (2l)}+\bar{\beta}e^{-ik(2l)}.
\end{align}
Note that we have taken this two-site periodicity, suitable for the diagonalisation of the Hamiltonian with the alternating chemical potential. Using this ansatz we obtain the following equations for the bulk spectrum
\new{
\begin{equation}
\begin{pmatrix}
\gamma_+ -2\epsilon(k)&\delta e^{-ik}+\delta^{-1}e^{ik}\\
\delta e^{ik}+\delta^{-1}e^{-ik}&\gamma_- -2\epsilon(k)
\end{pmatrix}
\begin{pmatrix}
\alpha\\ \beta
\end{pmatrix}=
\begin{pmatrix}
0 \\ 0
\end{pmatrix},
\end{equation}
}
which is satisfied if the determinant of the matrix vanishes, resulting in the following eigenvalues 
\new{
\begin{equation}\label{eq:pbcspectrum}
\epsilon_\text{bulk}^\pm(k)=\frac12\left(\mathcal{N}\pm\sqrt{q^2+q^{-2}+2\cos(2k)}\right).
\end{equation}
}
It is important to note that we have not yet specified anything for $k$. If the system were periodic, we would find $k=\frac{2\pi n}L$ for $n \in \{0,\ldots,\frac{L}2-1\}$.

In the open system to find a constraint on $k$ we have to study the boundary conditions. To do so we return to $C^2=BB^\intercal$, because it offers four boundary equations, which we need to set the four free parameters ($\alpha$,$\bar{\alpha}$,$\beta$,$\bar{\beta}$):
\new{
\begin{widetext}
\begin{align}
\begin{pmatrix}
\delta^{-2} +\gamma_1^2 -4\epsilon(k)^2 & (\gamma_1+\gamma_-)\delta^{-1} &1& 0\\
(\gamma_1+\gamma_-)\delta^{-1}&\delta^{-2}+ \delta^2+\gamma_-^2 -4\epsilon(k)^2 & (\gamma_++\gamma_-)\delta  &1\\
\end{pmatrix}&
\begin{pmatrix}
T(1)\\
T(2)\\
T(3)\\
T(4)\\
\end{pmatrix}=
\begin{pmatrix}
0 \\ 0
\end{pmatrix},\\
\begin{pmatrix}
1&(\gamma_++\gamma_-)\delta&\delta^{-2}+ \delta^2+\gamma_+^2 -4\epsilon(k)^2& (\gamma_++\gamma_L)\delta^{-1}\\
0&1&(\gamma_++\gamma_L)\delta^{-1}&\delta^{-2} +\gamma_L^2 -4\epsilon(k)^2
\end{pmatrix}&
\begin{pmatrix}
T(L-3)\\
T(L-2)\\
T(L-1)\\
T(L)\\
\end{pmatrix}=
\begin{pmatrix}
0 \\ 0
\end{pmatrix},
\end{align}
\end{widetext}
where we have dropped the super- ($a$) and subscript ($k$) for brevity. Plugging in the ansatz gives a $4\times4$ matrix. The determinant of this matrix vanishes when $\epsilon(k)=\epsilon_\text{bulk}^\pm(k)$ with $k=\frac{\pi n}L$ with $n \in \{1,\ldots,\frac{L}2-1\}$. Given this quantisation on $k$ we thus recover \eqref{eq:epsilonk}.} For $k=0$ the determinant also vanishes, but that occurs because the ansatz eigenfunction becomes trivial. Therefore, we can only determine $2(L/2-1)=L-2$ eigenvalues from $C$ directly. Adding the zero mode gives us $L-1$ eigenvalues. It turns out we can construct the remaining mode explicitly. The following Majorana modes satisfy $[H,T^{a,b}_{\mathcal{N}/2}]=\pm\frac{\mathcal{N}}2T^{a,b}_{\mathcal{N}/2}$:
\begin{align}
T^a_{\mathcal{N}/2}&=\alpha_1\sum_{j=1}^{L/2}\left(-q^{-2}\right)^{j-1}\left(a_{2j-1}+\frac{\eta}{q}a_{2j}\right),\label{eq:EMxeven}\\
T^b_{\mathcal{N}/2}&=\beta_1\sum_{j=1}^{L/2}\left(-q^{-2}\right)^{j-1}\left(b_{2j-1}+\frac1{q\eta}b_{2j}\right),\label{eq:EMyeven}
\end{align}
therefore the final eigenvalue is $\mathcal{N}/2$.

\section{Phase transition}\label{app:phasetransition}
Here we show that the phase transition at $\eta=1$ is in the DN-PT universality class. First we will address the non-interacting problem, and then also consider the interactions. We use the results for the staggered XXZ chain
\begin{align}
H=& \ -\frac{1}2\sum_{j=1}^L\bigg[\sigma_{j}^x\sigma_{j+1}^x+\sigma_{j}^y\sigma_{j+1}^y+\Delta \sigma_{j}^z\sigma_{j+1}^z\nonumber\\
&\qquad\qquad +2(h+(-1)^jh_s)\sigma_{j}^z\bigg].
\end{align}
In Ref.~\onlinecite{alcaraz1995critical} it was shown that there is a DN-PT phase transition at
\begin{equation}\label{eq:DNPTcondition}
h=\sqrt{h_s^2+1}-\Delta.
\end{equation}
For the non-interacting model with PBCs, then we can read from Eq.~\eqref{eq:lochamspin} that at criticality [$\eta=1$, recall Eq.~\eqref{eq:parameters}]
\begin{equation}
\left.
\begin{array}{rl}
h-h_s&=q\\
h+h_s&=q^{-1}
\end{array}
\right \}\Rightarrow
\begin{array}{rl}
h&=(q^{-1}+q)/2,\\
h_s&=(q^{-1}-q)/2.
\end{array}
\end{equation}
Thus with $J=1$ and $\Delta=0$ we see that Eq.~\eqref{eq:DNPTcondition} is satisfied, i.e., the model is precisely at the DN-PT transition. Similarly, when including interaction the phase transition still occurs at $\eta=1$. From Eqs.~(\ref{eq:XYZ}-\ref{eq:B0B1}) we can derive that
\begin{align}
h&=\cot\phi\ (q^{-1}+q),\nonumber\\
\Delta&=(\cot\phi-1)(q^{-1}+q),\\
h_s&=q^{-1}-q,\nonumber
\end{align}
where an overall factor of $\sqrt{2}\sin(\phi)$ has been taken out. Thus the condition \eqref{eq:DNPTcondition} is also satisfied for the interacting model.

\section{Ground state for PBCs and APBCs}\label{app:GSPBCAPBC}
Following App.~D in Ref.~\onlinecite{kawabata2017exact} we derive the ground states for PBCs and APBCs. From Eq.~\eqref{eq:factorizedGS} we define
\begin{equation}
A^\pm_L=\prod_{k=1}^{L/2}(x_0 c_{2k-1}\dagg\pm 1)(x_1 c_{2k}\dagg \pm1),
\end{equation}
such that $\ket{\Psi^\pm}=A^\pm\vac$. Subsequently
\begin{equation}
A^e_L=A^+_L+A^-_L,\qquad A^o_L=A^+_L-A^-_L,
\end{equation}
which implies $\ket{\Psi^{e,o}}=\frac1{\sqrt{2(N\pm M)}}A^{e,o}\vac$. 

We will now prove that $\ket{\Psi^o}$ $\left(\ket{\Psi^e}\right)$ is the ground state for PBCs (APBCs).
The open chain we studied in Sec.~\ref{sec:noninteracting} is closed by adding a boundary term, as we saw in Eq.~\eqref{eq:twistedboundary}. For PBCs and APBCs we can identify
\begin{equation}
h_{\rm bound}=h_L
\end{equation}
with $h_L$ as in Eq.~\eqref{eq:lochamfermion}, which acts on site $L$ and $L+1\equiv 1$. For PBCs we can identify $c_{L+1}=c_1$. Therefore, $h_{\rm bound}$ is minimised by $(x_1c_L\dagg\pm1)(x_0c_1\dagg\pm1)f(c_2\dagg,\ldots,c_{L-1}\dagg)$, where $f$ is some polynomial. Rewriting 
\begin{equation}
A^\pm_L=A^\pm_{L-1} x_1c_L\dagg \pm A^\pm_{L-1}=-x_1c_L\dagg A^\mp_{L-1}\pm A^\pm_{L-1}\label{eq:Apmrewrite},
\end{equation}
brings $c_L\dagg$ next to $c_1\dagg$. Next we note that 
\begin{align}
A^o_L&=(x_1c_L\dagg+1)A^+_{L-1}-(x_1c_L\dagg-1)A^-_{L-1}\\
&=(x_1c_L\dagg+1)(x_0c_1\dagg+1)(\ldots)\nonumber\\
&\qquad-(x_1c_L\dagg-1)(x_0c_1\dagg-1)(\ldots), \nonumber
\end{align}
with $(\ldots)$ some polynomial in $c_2\dagg,\ldots,c_{L-1}\dagg$. Therefore $A^o_L\vac$ minimises $h_{\rm bound}$, so the ground state for PBCs is $\ket{\Psi^o}$.

For APBCs $c_{L+1}\dagg=-c_1\dagg$, so $h_{\rm bound}$ is minimised by $(x_1c_L\dagg\pm1)(-x_0c_1\dagg\pm1)f(c_2\dagg,\ldots,c_{L-1}\dagg)$.  Using Eq.~\eqref{eq:Apmrewrite} we recognise
\begin{align}
A^e_L&=-(x_1c_L\dagg-1)A^+_{L-1}-(x_1c_L\dagg+1)A^-_{L-1}\\
&=-(x_1c_L\dagg-1)(x_0c_1\dagg+1)(\ldots)\nonumber\\
&\qquad-(x_1c_L\dagg+1)(x_0c_1\dagg-1)(\ldots), \nonumber
\end{align}
concluding that $\ket{\Psi^e}$ is the ground state for APBCs.

\section{Odd projector}\label{app:oddprojector}
In this appendix we derive the ground states of the Hamiltonian in Eq.~\eqref{eq:lochamint2}. It is known that there are $L+1$ unique ground states.\cite{cerezo2016factorization,cerezo2017factorization} For notational convenience we rewrite
\begin{align}
h_j^\text{int}=&-\mathcal{N}\Bigg[\cos(\theta_j)\left(\sigma_j^x\sigma_{j+1}^x+\sigma_{j}^y\sigma_{j+1}^y\right)+\sigma_{j}^z\sigma_{j+1}^z\nonumber\\
&+\sin(\theta_j)\left(\sigma_j^z-\sigma_{j+1}^z\right)-1\Bigg],\label{eq:lochamintapp}
\end{align}
where we used that 
\begin{equation*}
\cos(\theta_j)^2 +\sin(\theta_j)^2=\frac{(\eta+\eta^{-1})^2+(q_j-q_j^{-1})^2}{\mathcal{N}^2}=1.
\end{equation*}
One can easily check that the polarised state $\ket{\Uparrow}$ is one of the ground states of the system, $H\ket{\Uparrow}=0$. We claim that all the ground states are given by
\begin{equation}
\ket{\Psi_p}=(\tilde{\mathcal{S}}^-)^p\ket{\Uparrow},
\end{equation}
for $p=0,1,\ldots,L$ with $\tilde{\mathcal{S}}^-$ defined in Eq.~\eqref{eq:lowering}, such that $H\ket{\Psi_p}=0$. 

In order to prove this statement we need the following two conditions:
\begin{align}
H \tilde{\mathcal{S}}^- \ket{\Uparrow} = [H, \tilde{\mathcal{S}}^-] \ket{\Uparrow}&=0, \label{eq:cond2}
\\
[ [H, \tilde{\mathcal{S}}^-], \tilde{\mathcal{S}}^-] &= 0. \label{eq:cond3}
\end{align}
To derive this, we plug Eq.~\eqref{eq:lowering} into Eq.~\eqref{eq:cond2} resulting in the following constraint on $g_j$:
\begin{align}
g_j \cos(\theta_j)&=g_{j+1}(1-\sin(\theta_j)),
\end{align}
is satisfied by
\begin{equation}\label{eq:gjp1}
g_{j+1}=\frac{1+\sin(\theta_j)}{\cos(\theta_j)}g_j=\frac{\mathcal{N}+q_j-q_j^{-1}}{\eta+\eta^{-1}}g_j.
\end{equation}
Choosing $g_1=\sqrt{1-\frac{q-q^{-1}}{\mathcal{N}}}$ results in Eq.~\eqref{eq:lowering}, which proves Eq.~\eqref{eq:cond2}. Finally, simply writing out the commutator
\begin{equation}
[[h_j,g_j\sigma_{j}^z+g_{j+1}\sigma_{j+1}^z],g_j\sigma_{j}^z+g_{j+1}\sigma_{j+1}^z],
\end{equation}
and plugging in $g_j$ verifies Eq.~\eqref{eq:cond3}.

Now suppose $\ket{\Psi_p}$ and $\ket{\Psi_{p+1}}$ are ground states of $H$. For $p=0$ this is true, because of $H\ket{\Uparrow}=0$ and Eq.~\eqref{eq:cond2}. Using Eq.~\eqref{eq:cond3}:
\begin{align}
0&=[ [H, \tilde{\mathcal{S}}^-], \tilde{\mathcal{S}}^-]\ket{\Psi_p}\nonumber\\
&= \left(H (\tilde{\mathcal{S}}^-)^2-2\tilde{\mathcal{S}}^-H\tilde{\mathcal{S}}^-+(\tilde{\mathcal{S}}^-)^2H\right)\ket{\Psi_p}\nonumber\\
&=H\ket{\Psi_{p+2}}-2\tilde{\mathcal{S}}^-H\ket{\Psi_{p+1}}+(\tilde{\mathcal{S}}^-)^2H\ket{\Psi_p}\nonumber\\
&=H\ket{\Psi_{p+2}}.
\end{align}
Hence also $\ket{\Psi_{p+2}}$ is a zero-energy ground state of $H$, and therefore by induction all $\ket{\Psi_p}$ are ground states. Next, we note that $(\sigma_j^-)^2=0$, and therefore $(\tilde{\mathcal{S}}^-)^{L+1}=0$, allowing for finite $(\mathcal{S}^-)^p$ for $p=0,1,\ldots,L$.

Finally, we have to prove that (a) all $\ket{\Psi_p}$ exist and (b) we have found a complete ground-state basis.

\emph{(a):} For the first point we note that $g_j>0$ for all $j$, therefore $(\mathcal{S}^-)^p$ is a non-negative matrix and $(\mathcal{S}^-)^p\neq0$ for $p\leq L$. Hence, the states $\ket{\Psi_p}$ are non-trivial (i.e., $\norm{\ket{\Psi_{p}}}>0$).

\emph{(b):} For the second point we observe that $S^z\ket{\Psi_p}=(L/2-p)\ket{\Psi_p}$, with $S^z$ as in Eq.~\eqref{eq:Sz}. In other words, all $\ket{\Psi_p}$ belong to different $S^z$ sectors. Proving the completeness is equivalent to showing that the ground state $\ket{\Psi_{p}}$ is the unique ground state in the respective $S^z$ sector. Given that $S^z$ commutes with the Hamiltonian, the Hamiltonian matrix ($\mathcal{H}$) becomes block diagonal in the $S^z$ basis. Each block ($\mathcal{H}_p$) corresponds to a fixed $S^z$ sector. In Eq.~\eqref{eq:lochamintapp} we note that only the $\cos(\theta_j)$ terms gives rise to off-diagonal terms in $\mathcal{H}_p$, with a strictly negative coefficient $-(\eta+\eta^{-1})$.  Because $\sigma_j^x\sigma_{j+1}^x+\sigma_{j}^y\sigma_{j+1}^y$ corresponds to a non-negative matrix, the off-diagonal elements of the matrix are non-positive. Note that we are allowed to add a constant term, making the full matrix non-positive. Since $\mathcal{H}_p$ is hermitian, irreducible and non-positive, the Perron-Frobenius theorem tells us that the ground state is non-degenerate.\cite{berman1994nonnegative}$\hfill\blacksquare$

\end{document}